\documentclass[11pt, manuscript, nonacm]{acmart}

\renewcommand\footnotetextcopyrightpermission[1]{} 
\setcopyright{none}

\makeatletter
\let\@authorsaddresses\@empty
\makeatother

\usepackage{natbib}
\usepackage{amsthm,graphicx,multirow,amsmath,color,amsfonts}
\usepackage{subfig}
\usepackage{booktabs}
\usepackage{algorithm} 
\usepackage{algpseudocode}

\AtBeginDocument{%
  \providecommand\BibTeX{{%
    \normalfont B\kern-0.5em{\scshape i\kern-0.25em b}\kern-0.8em\TeX}}}

\acmConference[MobiHoc '21]{MobiHoc '21: ACM International Symposium on Mobile Ad Hoc Networking and Computing}{July 26--30, 2021}{Shanghai, China}
\acmBooktitle{Mobihoc '21: ACM International Symposium on Mobile Ad Hoc Networking and Computing,
  July 26--30, 2021, Shangai, China}
\acmPrice{15.00}
\acmISBN{978-1-4503-XXXX-X/18/06}

\begin{document}
\title{Improving Spectral Efficiency of Wireless Networks through Democratic Spectrum Sharing}

\author{Aniq Ur Rahman}
\orcid{0000-0003-3685-7201}
\email{aniqur.rahman@kaust.edu.sa}

\author{Mustafa A. Kishk}
\orcid{0000-0001-7518-2783}
\email{mustafa.kishk@kaust.edu.sa}

\author{Mohamed-Slim Alouini}
\orcid{0000-0003-4827-1793}
\email{slim.alouini@kaust.edu.sa}


\thanks{The authors are with King Abdullah University of Science and
Technology (KAUST), Thuwal 23955-6900, Saudi Arabia. E-mail: \{ \texttt{aniqur.rahman},
\texttt{mustafa.kishk}, \texttt{slim.alouini} \} \texttt{@kaust.edu.sa} }

\renewcommand{\shortauthors}{Rahman, Kishk and, Alouini}

\begin{abstract}
Wireless devices need spectrum to communicate. With the increase in the number of devices competing for the same spectrum, it has become nearly impossible to support the throughput requirements of all the devices through current spectrum sharing methods.
In this work, we look at the problem of spectrum resource contention fundamentally, taking inspiration from the principles of globalization. We develop a distributed algorithm whereby the wireless nodes democratically share the spectrum resources and improve their spectral efficiency and throughput without additional power or spectrum resources.
We validate the performance of our proposed democratic spectrum sharing (DSS) algorithm over real-world Wi-Fi networks and on synthetically generated networks with varying design parameters. Compared to the greedy approach, DSS achieves significant gains in throughput ($\sim$60\%), area spectral efficiency ($\sim$50\%) and fairness in datarate distribution ($\sim$20\%). Due to the distributed nature of the proposed algorithm, we can apply it to wireless networks of any size and density. 
\end{abstract}

\maketitle
\thispagestyle{empty}


\textcolor{black}{Connecting a massive number of wireless devices to the internet with a high datarate ~\cite{ilderem2020} depends on the availability of more spectrum resources. In 5G, we saw the addition of millimeter-wave spectrum to meet this demand. Similarly, in the upcoming generation of wireless systems, called 6G, we anticipate the use of frequencies in the terahertz (THz) range \cite{nagatsuma2016advances, koenig2013wireless, ghasempour2020single, ma2017frequency, papasotiriou2021}. However these high frequency signals suffer more attenuation compared to their sub-6GHz counterpart, which makes the coverage size of a single transmitter prohibitively small~\cite{dang2020should}. To make up for the signal power lost to attenuation, the transmitters can increase their power. Alternatively, the network could be made denser by adding more transmitters. Unfortunately, both of these methods consume more energy and therefore, is not environment-friendly~\cite{chih2020energy}.}
Looking at the famous equation given by Claude E. Shannon for channel capacity \cite{shannon1949}, one can suggest two ways to improve the throughput of the wireless channel: one by increasing the channel bandwidth and the other by decreasing the interference. The first option requires more resources, i.e., bandwidth, while the second option improves the spectral efficiency without demanding additional resources.

For example, consider this common problem in wireless networks:
when nearby access points (APs) utilize the same spectrum resources, they degrade each other's performance by causing mutual interference. Since the APs are constrained by design and regulation to use certain spectrum bands of finite bandwidth, the only solution is to manage the interference cooperatively.
In order to manage interference, the nodes can adjust their transmission power \cite{zhang2020deep, zhang2017qoe, zhang2016learning}, or they can choose parts of the spectrum to transmit the signals over \cite{tan2021cooperative, liang2019spectrum, zhang2017qoe, naparstek2018deep, chen2020multi, chen2013database, zhang2016learning}. In this paper, we are more interested in the latter technique, which involves allocating spectrum resources among the wireless nodes to improve the throughput distribution among them collectively. The motivation is primarily due to the fact that spectrum resources are limited, and more so, in the crowded lower bands, which are desirable because of their propagation characteristics. Moreover, as the user demands increase with time, efficient utilization of spectrum resources becomes more critical.

\textcolor{black}{Historically, spectrum sharing among different technologies have been carried out on a time-sharing basis~\cite{abbott1999}, which proportionally reduces the throughput of each technology involved. This primitive solution to spectrum resource contention leads to spectrum wastage, and there exists potential to achieve better spectral efficiency through demand-based consensus. Intelligent spectrum sharing algorithms could therefore adapt to the changes in user demands and spectrum availability to maximize the spectral efficiency~\cite{medard20205}.}

The spectrum allowed to be used by the wireless nodes is divided into a set of sub-bands. Since the nearby nodes cause mutual interference, one way to mitigate interference can be to encourage neighboring nodes to use a mutually exclusive set of sub-bands. However, this process is not realistic, as the nodes will not be willing to sacrifice sub-bands if their throughput requirements are not met. Therefore, we design a framework where each node takes into account the suggestions of its neighboring nodes along with its own throughput requirements, while deciding on sub-band occupation. 
We take inspiration from the concept of globalization which promotes cooperation among nation-states to improve cross-border trade and facilitate the free exchange of goods, services, cultures, and ideas. According to Steger \cite{steger2003}, the major ideological claims of globalization are: (i) it liberates and integrates markets around the world, (ii) there is no hierarchy in decision making, (iii) it benefits all the parties involved, and (iv) it promotes democracy. In the context of spectrum sharing, globalization can promote cooperation among neighboring nodes to take decisions in a distributed manner. The nodes are allowed to be selfish to a certain degree to safeguard the interests of the users it serves. Moreover, the results should exhibit a fair distribution of spectrum resources among the nodes. The following analogy can be drawn from globalization for modeling the interaction of the APs. The AP or node acts as a nation; the user equipments that each AP serves can be viewed as citizens, and the finite resource being shared here is the spectrum. Consequently, we name our proposed scheme democratic spectrum sharing (DSS) as the neighboring nodes have a say in the decision-making process.

The works on distributed spectrum sharing \cite{tan2021cooperative, gao2021hermes, chen2020multi, zhang2020deep, moy2020decentralized, liang2019spectrum, naparstek2018deep, zhang2017qoe, zhang2016learning, chen2013database, cao2005distributed} use a variety of mathematical frameworks like game theory, multi-agent reinforcement learning (MARL), deep reinforcement learning (DRL) and multi-armed bandits (MAB). Even though the topic of distributed spectrum sharing is well-studied, the majority of the literature focuses on a hierarchical setup of cognitive radios, where the secondary users (SUs) share the spectrum with the primary users (PUs).
Some articles which discuss non-hierarchical spectrum sharing lack a social component, as all the nodes are modeled as selfish agents that try to improve their own performance \cite{chen2020multi}. These agents take actions based on the observations from the environment without getting direct feedback from the neighboring APs. Furthermore, some works \cite{naparstek2018deep, chen2020multi} consider only a single sub-band being shared among all the APs, which makes the problem unrealistic since in practice, a channel is composed of multiple sub-bands.
Our work bridges these gaps as we focus on the unlicensed spectrum where all the APs have equal access rights. The APs cooperate with the neighboring APs to transmit over a subset of multiple sub-bands to mitigate the effect of interference and hence achieve better spectral efficiency and throughput.

\begin{table*}[ht!]
    \centering
    \caption{Distributed spectrum sharing techniques.}
    \begin{tabular}{  p{4 cm}  p{1 cm}  p{3.5 cm}  p{3.5 cm}  }
    
    \toprule
    
    \textbf{Reference} & \textbf{Year} &  \textbf{Mathematical \newline framework} & \textbf{Agents/Players} \\ \midrule 
    
    Tan \textit{et. al.} \cite{tan2021cooperative} & 2021 &  Game theory \& MARL & User equipments \\ \hline
    
    Gao \textit{et. al.} \cite{gao2021hermes} & 2021 & Improved MARL & User equipments \\ \hline
    
    Chen and Vasal \cite{chen2020multi} & 2020 & MARL & Base stations \\ \hline
    
    Zhang and Liang \cite{zhang2020deep} & 2020 &  DRL & Access points \\ \hline
    
    Moy \textit{et. al.} \cite{moy2020decentralized} & 2020 & MAB & IoT devices \\ \hline
    
    Liang \textit{et. al.} \cite{liang2019spectrum} & 2019 & MARL & Vehicles \\ \hline
    
    Naparstek and Cohen \cite{naparstek2018deep} & 2019 &  Game theory \& MARL & User equipments \\ \hline
    
    Zhang \textit{et. al.} \cite{zhang2017qoe} & 2017 &  Game theory & Base stations \\ \hline
    
    Zhang \textit{et. al.} \cite{zhang2016learning} &  2016 & MAB \& graph-coloring & Access points \\ \hline
    
    Chen and Huang \cite{chen2013database} & 2013 & Game theory & TVWS access points \\ \hline
    
    Cao and Zheng \cite{cao2005distributed} & 2005 &  Game theory & Access points \\ 
    \bottomrule
    
    \end{tabular}
    
    \label{tab:relworks}
\end{table*}

In centralized spectrum sharing, it is possible to arrive at the optimal solution by solving an optimization problem. However, a centralized architecture enforces hierarchy in decision making and requires all the nodes to communicate with the central controller. Moreover, the time taken to find the optimal solution increases as the network size grows, suggesting that centralized spectrum sharing is not scalable. In distributed spectrum sharing, we can achieve near-optimal results in much lesser time compared to centralized approach. In DSS, we tread the middle path, where we get cooperation of a centralized approach, and scalability of a distributed system.

\section*{Results}

\subsection*{Democratic Spectrum Sharing}

We consider a network of $N$ APs operating in the unlicensed spectrum which has a total of $S$ sub-bands, each of the same bandwidth $W$. Each AP secures the sub-bands for the users within its coverage radius $R$ and transmits their signals with power $P_T$. The available bandwidth can then be allocated for downlink and uplink transmissions. In this paper, we only focus on the downlink scenario considering the datarate QoS.
Taking inspiration from belief propagation in graphs~\cite{frey1998revolution}, we first construct an interference graph of the nodes, where the neighboring nodes that are close enough to potentially cause interference are connected. This distance is named the neighborhood radius, and is denoted by $R_N$. Each node tries to avoid using the sub-bands occupied by its neighbors akin to a multi-graph coloring approach. We call this the social decision which is the result of a voting process where the occupancy of a sub-band by a neighboring node is weighted by a factor related to the average path-loss. If the social decision does not allow the node to meet its throughput requirements, it exploits more sub-bands if available. We refer to this as the selfish decision. Each AP is equipped with an independent Poisson clock \cite{rahman2021game}, and the sub-band occupation algorithm runs every time the Poisson clock is triggered. This allows for a truly decentralized operation without the need of synchronization.

\begin{figure}[h]
    \centering
    \includegraphics[width=5.9in]{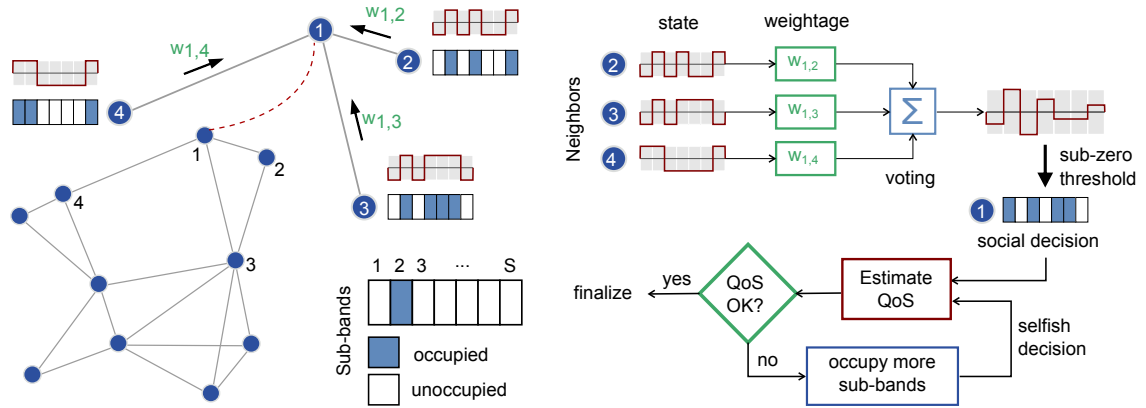}
    \caption{\textbf{Democratic spectrum sharing mechanism.} The figure shows a sample network and demonstrates how node 1 takes a decision regarding sub-band allocation after voting from the neighboring nodes 2,3 and 4.}
    \label{fig:mech}
\end{figure}

In Fig.~\ref{fig:mech} we demonstrate the mechanism with the help of four nodes in the network. Here, node~1 gets the sub-band occupation state (SBOS) of its neighboring nodes $\{2,3,4\}$, following which a voting takes place leading to the social decision. Then node 1 estimates the QoS that it can achieve by taking the social decision. The mechanism then allows node 1 to occupy more sub-bands until its QoS requirements are met. This feedback loop leads to a selfish decision.


\subsection*{Analysis of Synthetic Wi-Fi Networks}
We first test our proposed framework on various synthetic network deployments. This enables us to study the framework in detail and understand how different parameters such as node density $\lambda$, neighborhood radius $R_N$ and number of nodes in the network $N$ affect the performance of DSS.
Throughout this paper, we compare the performance of DSS with the greedy approach wherein each node exploits all the sub-bands without any cooperation with its neighbors.

\begin{figure}[h]
    \centering
    \includegraphics[width=5.9in]{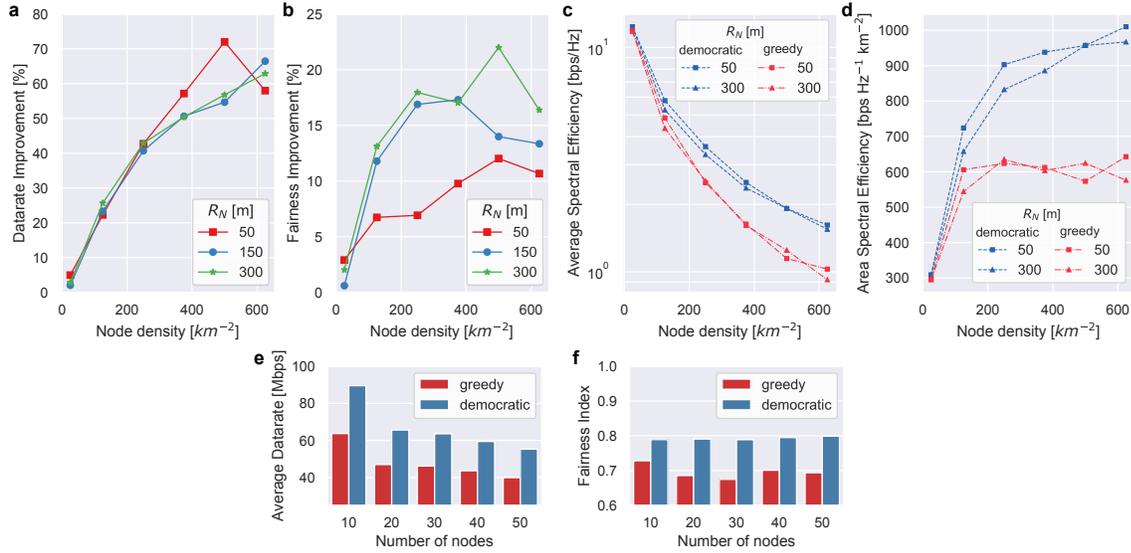}
    \caption{\textbf{Performance improvement in synthetic networks.} \textbf{a},\textbf{b}, The improvement in datarate and fairness index is shown in the figures (a) and (b) respectively, as the node deployment density $\lambda$ and the neighborhood radius $R_N$ are varied across a range of values. \textbf{c}, Average spectral efficiency vs. node density for different $R_N$. \textbf{d}, Area spectral efficiency vs. node density with varying $R_N$. \textbf{e}, Average datarate for different number of nodes. \textbf{f}, Fairness index for different number of nodes. }
    \label{fig:df_synthetic}
\end{figure}

In Fig.~\ref{fig:df_synthetic} we showcase the performance gains provided by DSS in terms of the datarate and fairness index \cite{jainfair}. In Fig.~\ref{fig:df_synthetic}.a, the datarate improvement is more significant for denser networks. For $\lambda \geq 375$ km$^{-2}$, we see that the improvement is more for smaller $R_N$, except at $\lambda=625$ km$^{-2}$, where $R_N=150$~m performs the best, and $R_N =50$~m the worst. This could indicate that as networks become denser, we must cover enough neighbors to get good feedback.

In Fig.~\ref{fig:df_synthetic}b, we study the improvement in fairness index as the network becomes denser for different values of $R_N$. For $\lambda=25$ km$^{-2}$, the improvement is insignificant, however, for dense networks, the improvement in fairness is more pronounced. We also see that the fairness index improves as $R_N$ increases, the only anomaly being sparse networks. In sparse networks, if $R_N$ is large, the node will receive feedback from a neighbor far away which does not cause much interference in the first place.
The spectral efficiency achieved by the democratic and greedy techniques is compared in Fig.~\ref{fig:df_synthetic}.c, where DSS outperforms the greedy approach for any network density, the difference mostly highlighted for dense networks. Similarly, in Fig.~\ref{fig:df_synthetic}.d, we show the area spectral efficiency \cite{alouini1999area} where the difference in the performance of DSS and greedy approach are almost negligible for sparse networks.

Finally, in Fig.~\ref{fig:df_synthetic}.e and Fig.~\ref{fig:df_synthetic}.f, we plot the average datarate and fairness index respectively, as the number of nodes in the network increases for the same node density. The average datarate decreases with increase in network size, however the gap between DSS and greedy approach remains the same. When it comes to fairness index, we get consistent values $\sim 0.8$ for networks of any size.

\subsection*{Analysis of Wi-Fi Networks in Glasgow}
To test the performance of our proposed DSS algorithm, we analyze the Wi-Fi AP distribution in the city of Glasgow, United Kingdom. The map consists of a total of 19,124 APs scattered in an area of 972 $\rm{km}^2$. We divide the entire map into $50 \times 50$ grids, and then evaluate the performance of DSS on each grid separately. The values of the parameters are: $S = 10$, $W = 20$~MHz, $P_T= 1$~W, $R= 30$~m, and $R_N=300$~m.

\begin{figure}[h]
    \centering
    \includegraphics[width=5.9in]{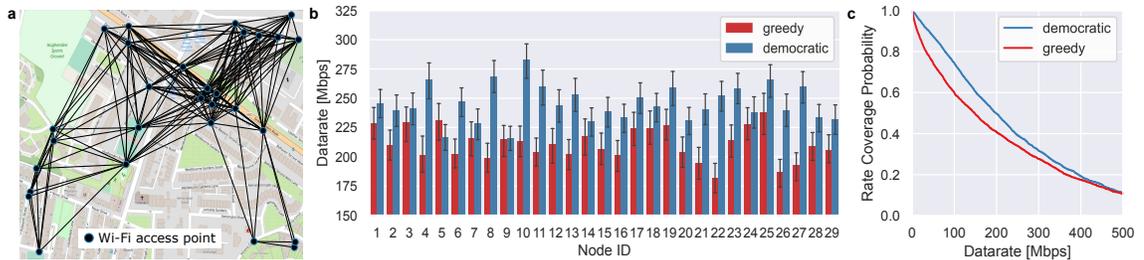}
    \caption{\textbf{Democratic spectrum sharing in action.} \textbf{a}, A part of Glasgow, UK is shown with 29 Wi-Fi APs that are depicted as black nodes in the map. The nodes that are closer than 300 metre are connected through black lines indicating that they are neighbors. \textbf{b}, The datarate of all the nodes in the network as a result of DSS and greedy schemes is shown through a box plot. \textbf{c}, Rate coverage probability or complementary cumulative density function of the datarates.}
    \label{fig:sample}
\end{figure}

We first show a sample grid consisting of 29 APs in Fig.~\ref{fig:sample} and construct an interference graph with neighborhood radius of 300 m. The datarates achieved by the nodes by greedy method are set as minimum datarate thresholds in the DSS algorithm. We evidently see that the average datarates in DSS (\ref{fig:sample}.b) are better than greedy approach (\ref{fig:sample}.c). We also calculate the fairness index of the datarate distribution among the nodes: compared to $0.84$ of the greedy method, DSS achieves a fairness index of $0.92$.

\begin{figure}[h]
    \centering
    \includegraphics[width=5.9in]{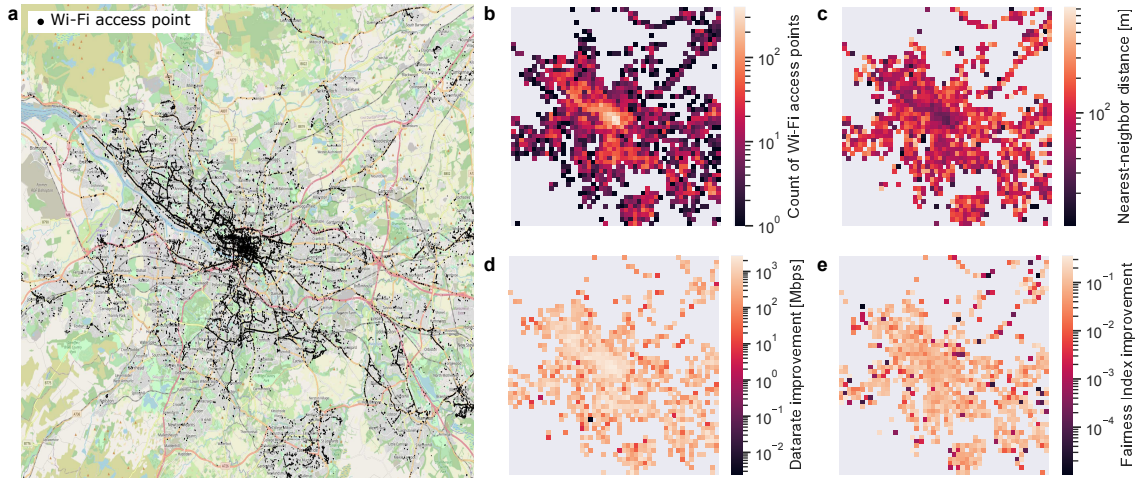}
    \caption{\textbf{Grid-wise analysis of the Wi-Fi networks in Glasgow.} \textbf{a}, The entire city of Glasgow, UK is shown with the APs represented by black dots. The map is divided into $50 \times 50$ grids of same area which are analyzed individually. \textbf{b}, The number of APs in each grid is conveyed through a heat map. \textbf{c}, For each grid, we record the nearest neighbor distance for all the nodes and present the average value in the heat map. \textbf{d, e}, The improvements in datarate and fairness index achieved as a result of DSS is shown through two heat maps. }
    \label{fig:map}
\end{figure}

In Fig.~\ref{fig:map}, we present the datarate and fairness improvements across all the networks in the 2500 grids on the Glasgow map. The node density is maximum towards the centre of the city (\ref{fig:map}.b), and the node distribution becomes sparse and clustered as one moves away from the centre. This fact is also highlighted by the average nearest-neighbor distance depicted in (\ref{fig:map}.c). In (\ref{fig:map}.d) and (\ref{fig:map}.e) we show the improvement in datarate and fairness index in each of the grids.

\subsection*{Analysis of Campus Wi-Fi Network at KAUST}
\begin{figure}[h]
    \centering
    \includegraphics[width=5.9in]{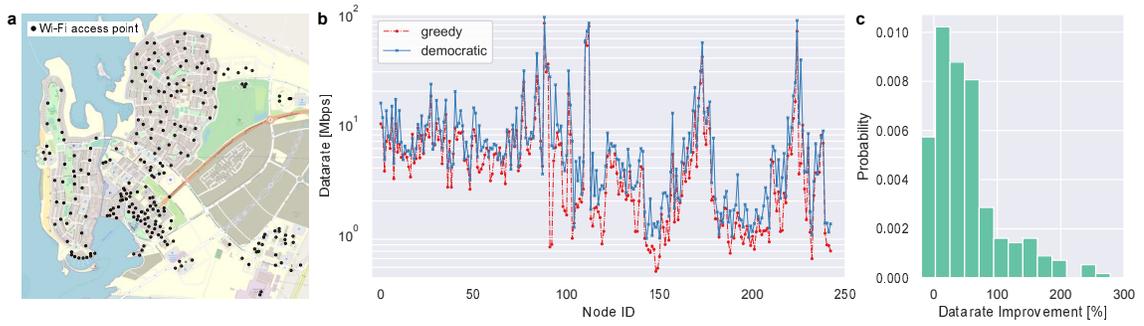}
    \caption{\textbf{Democratic spectrum sharing applied to a campus network.} \textbf{a}, A map of KAUST, Thuwal is shown with 243 Wi-Fi APs that are depicted as black nodes. \textbf{b}, The datarate of all the nodes in the network as a result of DSS and greedy schemes is shown through a line plot. \textbf{c}, The \% improvement in datarate is depicted through a normalized histogram.}
    \label{fig:kaust}
\end{figure}

After having shown the applicability of DSS in a large-scale city network, we now test the algorithm and evaluate its performance on a much smaller Wi-Fi network. For this, we consider the campus Wi-Fi network of King Abdullah University of Science and Technology (KAUST), in Thuwal, Saudi Arabia which consists of 243 access points spread over an area of 30 km$^2$. The results in Fig. \ref{fig:kaust} demonstrate considerable datarate improvements. Moreover, DSS improves fairness index from 0.25 to 0.32.

\section*{Discussion}

The DSS algorithm performs exceptionally well in dense networks, where it is needed due to the high amount of interference that the neighbors can cause due to proximity. DSS also achieves better area spectral efficiency than the greedy approach for all types of network deployment. Moreover, the performance improvement in terms of datarate and fairness is consistent for all network sizes, strongly suggesting that the algorithm is scalable.
The various opportunities and challenges associated with the democratic framework are discussed in the paragraphs that follow.

\subsubsection*{Non-Uniform Characteristics of the Sub-Bands:}
The propagation characteristics vary across different frequency bands, and the \textit{beachfront} spectrum, i.e., sub-6 GHz spectrum, is always in high demand due to its resilience to blockages. Therefore, the social decision should be unbiased, and the sub-bands belonging to the beachfront spectrum should not always be occupied by the same set of nodes and must be shared and utilized by all nodes on a time-sharing basis. 
Moreover, the edges between any two nodes are weighted by the average path-loss. Since the probability of having a line-of-sight connection also changes with frequency, the path-loss exponent will also depend on the sub-band in consideration, which must be accounted for in the democratic framework.

\subsubsection*{Presence of Rogue Nodes:}
Sometimes, there may be isolated nodes co-located within the network. These nodes are called rogue nodes, as they do not participate in the social decisions and do not send or receive any information regarding sub-band occupation. The interference from such nodes can be sensed and then used to update the noise factor $n_0$, which allows us to use the proposed framework as usual. The authors in \cite{sahai2008cognitive} suggest that for cognitive radios to be successfully deployed, the rogue nodes must be taken care of, i.e., every node that fails to behave well must be penalized.

\subsubsection*{Exchanging Information for Coordination:}
To make well-informed social decisions, the neighboring nodes need to send and receive information regarding their sub-band occupation state (SBOS). This communication overhead can be met by peer-to-peer communication over a common control channel, where a node, on updating its SBOS, sends this information to all its neighbors wirelessly to save and later use in their upcoming social decisions.
Another technique is that the node publishes its SBOS information to an online database accessible  \cite{chen2013database}  by all its neighbors. It does not require a common control channel and is, therefore, more reliable. However, there are delays associated with writing to and fetching data from an online database.

\subsubsection*{Exploiting White Spaces:}
There are instances where licensed users are not using portions of the licensed spectrum in a given geographical location. The unused portion of such spectrum is referred to as white spaces \cite{chen2013database}, and the nodes which do not own the license of such bands can then use them opportunistically. This is the basic premise of cognitive radios and can be easily incorporated in the democratic framework with a condition, i.e., the node should determine whether the sub-band belonging to white space is not in use through database queries or sensing methods. Then, the available sub-bands can be exploited just like the unlicensed spectrum bands.

\subsubsection*{Determining QoS Requirements:}
Each node first needs to determine the overall QoS requirement of the users under its coverage zone. This can be communicated by the applications being run on the user equipments. The idea is that a node with lesser QoS requirements will use lesser spectrum resources, which can then be used by the node with a higher QoS demand. Ensuring a fair distribution of spectrum resources in proportion to the demands of the nodes is a challenging problem, and more so in a distributed system.

\subsubsection*{Non-Stationary Environment:}
As the users are mobile and can roam from the coverage zone of one node to the other, the QoS requirements of each node will keep evolving in space and time, because of which, we must design a distributed algorithm that can adapt well to the changes. A famous technique is using multi-agent reinforcement learning and casting the social decision-making problem in the multi-armed bandit framework as done in \cite{moy2020decentralized, zhang2016learning}. For performance assessment, a realistic non-stationary environment must be designed, which would allow us to test how quickly the system adapts to the changes and reaches near-optimal performance.

\subsubsection*{Decision Trigger and Execution Time:}
The Poisson clock of each node is independent of other nodes' clocks. This decreases the probability of execution time overlap among neighboring nodes, which is essential for the following reason; If two or more neighboring nodes are in the midst of the decision-making process simultaneously, none of them would have the correct QoS estimates, which is essential for the algorithm. A simple fix is to use a semaphore, locked when a node is triggered and unlocked as soon as it publishes the updated information. At each trigger, the node must check whether the semaphores of all the neighboring nodes are unlocked before executing the decision-making algorithm.

\subsubsection*{Balance between Social and Selfish Decisions:}
The democratic framework is composed of two major decision-making blocks: social and selfish. Instead of having a fixed flow from social to selfish, we can, on each trigger, select either social or selfish decision block in a probabilistic manner. This social-selfish dilemma is similar to the explore-exploit dilemma of bandit algorithms and is an open question that deserves attention.

\section*{Methods}

Consider a weighted graph $\mathcal{G}(\mathcal{V}, \mathcal{E})$ where $\mathcal{V}$ is the set of APs and $\mathcal{E}$ is the set of edges connecting them. The distance between the APs $v$ and $v'$ is denoted by $d_{v,v'}$ and the weight of the edge connecting them is $w_{v, v'}$. The neighbors of AP $v \in \mathcal{V}$ are represented by the set $\mathcal{N}_v$. 
All the APs have a Poisson clock, i.e., the clock triggers follow a Poisson distribution with some parameter $\Lambda$. Let the sub-band occupation state (SBOS) of AP $v$ be denoted by the vector $\mathbf{b}_v \in \{-1,1\}^{S \times 1}$ where $S$ is the total number of sub-bands. An occupied sub-band has state 1 while an unoccupied sub-band has a state -1.
Let the SBOS suggested by the neighbors of AP $v$ through the voting mechanism be denoted as $\hat{\mathbf{b}}_v \in \mathbb{R}^{S \times 1}$. The algorithm for democratic spectrum sharing is presented in Algorithm~\ref{DSS}.

\begin{algorithm}[h!]
\caption{Democratic Spectrum Sharing}
\begin{algorithmic}[1]
    \If{Trigger} \Comment{The Poisson clock triggers}
    
    \State $\hat{\mathbf{b}}_v \gets \sum_{v' \in \mathcal{N}_v} w_{v,v'} \mathbf{b}_{v'}$ \Comment{The neighbors vote}

    \State $\mathbf{b}_v[k] \gets - {\rm sgn} ( \hat{\mathbf{b}}_v[k] ) , \forall k \in [1,S]$ \Comment{\textit{Social decision}: occupy sub-bands with negative vote}
    
    \State Estimate QoS with $\mathbf{b}_v$
    
    \While{the QoS requirement is not satisfied \& all the sub-bands are not occupied}
    
    \State $k \gets \arg \min \hat{\mathbf{b}}_v : \hat{\mathbf{b}}_v [k] > 0 $
    \Comment{occupy the sub-band with least positive vote}
    
    \State $\mathbf{b}_v[k] \gets 1$
    \Comment{\textit{Selfish decision}}
    \State $\hat{\mathbf{b}}_v[k] \gets -1$ \Comment{Mark it occupied}
    
    \State Estimate QoS with $\mathbf{b}_v$
    
    \EndWhile
    
    \EndIf
    
\end{algorithmic}
\label{DSS}
\end{algorithm}



\section*{Data Availability}
The dataset of access point locations in the city of Glasgow, UK is available at \cite{glasgowcitycouncil2014}. The dataset of KAUST Campus Wi-Fi network is available at \cite{dssgithub}.
The results generated during analysis is available at \cite{dssgithub}.

\section*{Code Availability}
A Python implementation of the algorithm, and the scripts used for analysis are available at \cite{dssgithub}.

\bibliographystyle{naturemag}
\bibliography{main.bib}

\end{document}